\documentclass[a4paper,11pt]{article}
\usepackage{pos}

\usepackage[utf8]{inputenc}

\title{Status of the NEXT experiment for neutrinoless double beta decay searches}

\author{Carmen Romo-Luque for the NEXT Collaboration}

\affiliation{Instituto de Física Corpuscular (IFIC), CSIC $\&$ Universitat de València,\\
  Calle Catedrático José Beltrán, 2, 46980 Paterna, Valencia, Spain}

\emailAdd{Carmen.Romo@ific.uv.es}

\abstract{NEXT (Neutrino Experiment with a Xenon TPC) is a neutrinoless double beta decay experiment located at the Laboratorio Subterráneo de Canfranc (LSC, Spain). Its aim is to demonstrate that the neutrino is a Majorana particle by detecting the neutrinoless double beta decay process in xenon gas enriched in the $^{136}$Xe isotope. The detector technology used in NEXT is that of radiopure high pressure time projection chambers with electroluminescence amplification, which provide excellent energy resolution better than 1$\%$ FWHM in the energy region of interest, topological reconstruction that allows rejecting single-electron background events and a strong potential for “in situ” tagging of the barium daughter ion. The experiment has been developing in phases.
	
The NEXT-White detector has recently finished operation at the LSC and contained approximately an active Xe mass of 5 kg. Its purpose was to demonstrate the excellent energy resolution, to validate the reconstruction algorithms and the background model, and to make a measurement of the two-neutrino double beta decay of $^{136}$Xe.
	
The 100 kg NEXT-100 detector is under construction and is scheduled to be installed and assembled by the first half of 2022. The predicted 90$\%$ CL sensitivity to the neutrinoless double beta decay half-life will reach $10^{26}$ years for an exposure of about 400 kg $\cdot$ year.
	
A vigorous program towards the development of ton-scale detectors is also under way, including extensive R$\&$D towards the realization of in-situ Ba$^{2+}$ tagging as means to achieve virtually zero-background detection. A first module with a mass of at least 500 kg may be operating as early as 2026 at the LSC.
	
In this manuscript, recent results obtained with the NEXT-White detector will be presented, as well as the NEXT-100 construction status and on the prospects of future NEXT detectors.}

\FullConference{%
  *** The 22nd International Workshop on Neutrinos from Accelerators (NuFact2021) ***\\
  *** 6–11 Sep 2021 ***\\
  *** Cagliari, Italy ***}

\begin{document}
\maketitle

\section{The NEXT experiment}

The NEXT experiment is searching for the extremely rare $0\nu\beta\beta$ decay of $^{136}$Xe using the technology of high pressure xenon gas (HPXe) time projection chamber (TPC) with electroluminescent amplification (EL). The detection of this process would imply that the neutrino is a Majorana particle and the total lepton number is not conserved, which could be related to the cosmological asymmetry between matter and antimatter through a mechanism known as leptogenesis \cite{NEXT:2015wlq} \cite{Fukugita:1986hr}. 

\smallskip

When a double beta decay occurs, two electrons are emitted and interact with the Xe gas through ionization and excitation. First, they excite xenon atoms, causing the emission of scintillation light detected by an array of PMTs located behind the cathode in a time t$_0$, and they ionize the gas and take off electrons of the last layer of this. The electrons produced in the ionization process are drifted by an electric field towards the anode, reaching the EL zone, where the electric field is more intense. There, the electrons are accelerated emitting around thousand  photo-electrons each, being detected again by the PMTs at a time t$_1$. The difference between the two times t$_0$ - t$_1$ provides the z position of the track of electrons. The forward-moving EL  photo-electrons are detected by a dense array of SiPMs located behind the anode, where the signals are used for track reconstruction.

\smallskip

The NEXT program started in 2009 with NEXT-DEMO (IFIC, Valencia, Spain) and NEXT-DBDB (Berkeley, California, USA), two prototypes holding 1 kg of xenon. They demonstrated the robustness of the technology, the excellent energy resolution and its unique topological signal.

\section{NEXT-White results}

NEXT-White \cite{NEXT:2018rgj} represents the first radiopure, large scale demonstrator of the NEXT Experiment. The detector contained 5 kg of xenon and had stable data taking from autumn 2016 to summer 2021 at the LSC in Canfranc. Its main results include the validation of the technology in a large-scale radiopure detector, background model assessment, reach an excellent energy resolution and efficient topology discrimination between single and double-electron tracks and measure the two-neutrino double beta decay of $^{136}$Xe.

\smallskip

\begin{figure}[h!]
	\centering
	\includegraphics[height=3.8cm]{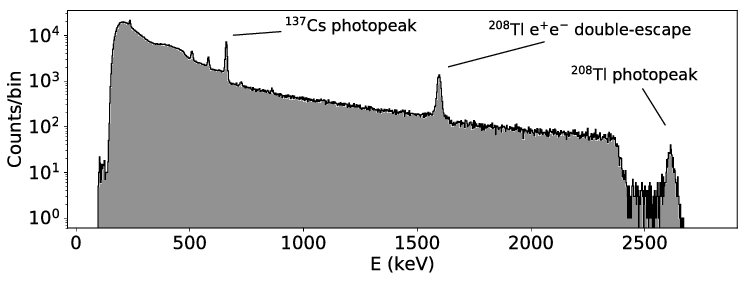}
	\includegraphics[height=3.8cm]{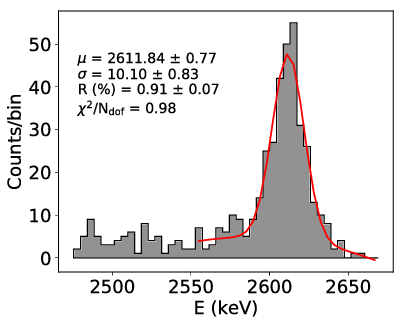}
	\caption{Left: full energy spectrum from calibration data with $^{228}$Th and $^{137}$Cs sources. Right:  energy spectra of the $^{228}$Th energy peak.}
	\label{fig:energy_spec}
\end{figure}

The energy resolution was measured at different energies using cesium and thorium calibration sources and reached 0.91$\%$ at 2.6 MeV. Figure \ref{fig:energy_spec} left shows the energy spectrum measured by the PMTs coming from the scintillation light from the EL region while Figure \ref{fig:energy_spec} right shows the fit of the $^{228}$Th peak \cite{NEXT:2019qbo}.

A plane of 1792 SiPMs was used to reconstruct the tracks of the particles within the detector. The signature of a $0\nu\beta\beta$ event in high pressure xenon gas consists of a single long track of constant energy deposition with two larger energy depositions at the end points (‘blobs’). This behaviour is unique to double beta decay and can be used to discriminate signal (double electron events with two end-points of higher energy) from background (mainly single electron and typically more than one track events with only one blob at one extreme).

\smallskip

\begin{figure}[!htb]
	\centering
	\includegraphics[height=5cm]{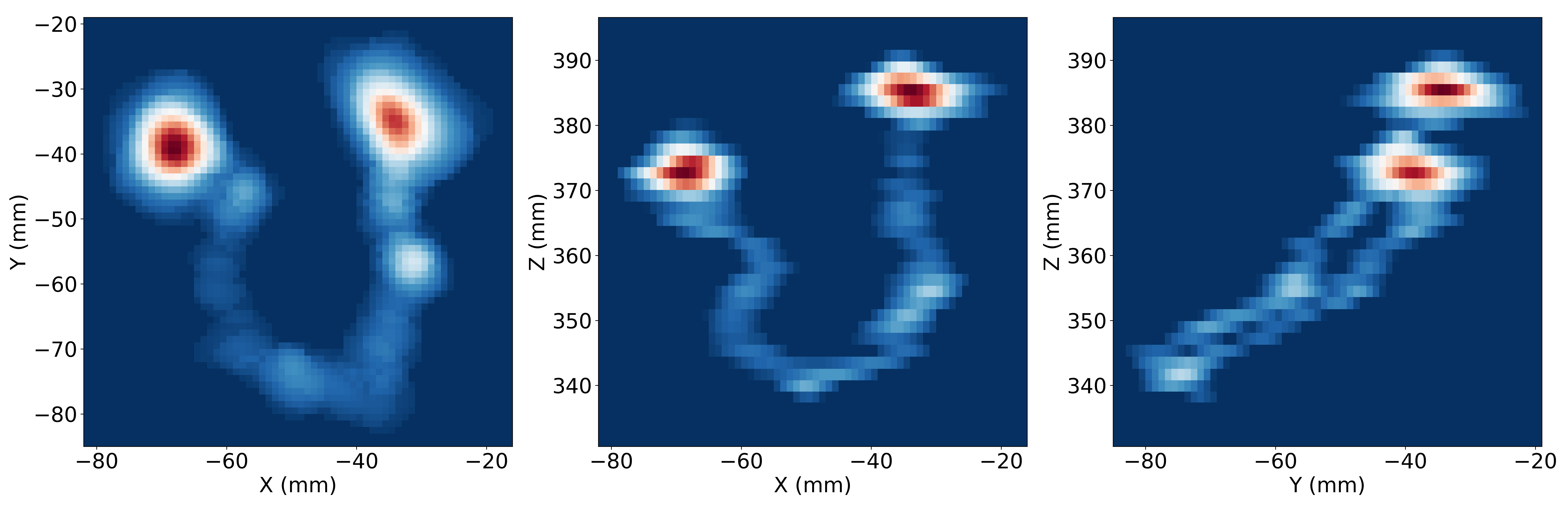}
	\caption{Deconvolved 2 MeV $2\nu\beta\beta$ candidate using the Richardson-Lucy algorithm obtained during the data taking of NEXT-White \cite{NEXT:2020try}.}
	\label{fig:Lucy_Rich_deconv}
\end{figure}

Once the 3D hits were reconstructed and corrected by electron drift lifetime, geometrical effects and time variations according to the data using the calibration source $^{83m}$Kr, a novel reconstruction through Richardson-Lucy iterative deconvolution was performed to reduce the blurring induced by the electron diffusion and EL light production. It provided a high definition image of the tracks, increasing the background rejection factor and the signal efficiency \cite{NEXT:2020try} \cite{NEXT:2021dqj}. Figure \ref{fig:Lucy_Rich_deconv} right shows the topology of a $2\nu\beta\beta$ candidate from data that was studied in \cite{NEXT:2020try} via the Richardson-Lucy algorithm.

\begin{figure}[!htb]
	\centering
	\includegraphics[height=5.5cm]{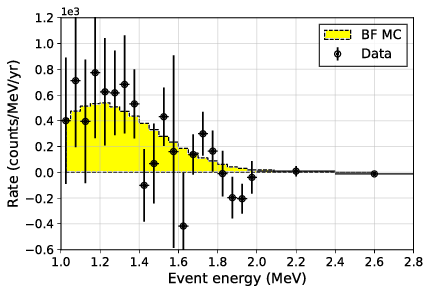}
	\includegraphics[height=5.3cm]{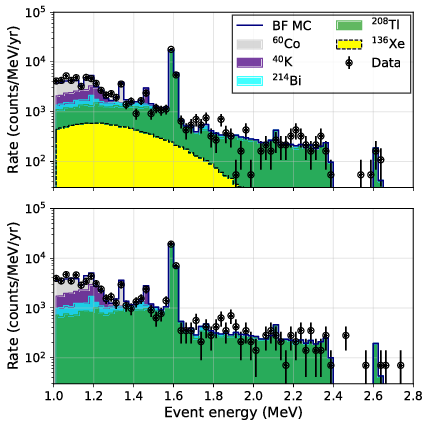}
	\caption{Left: Background-subtraction $2\nu\beta\beta$  fit. The background subtracted data (black dots) with its statistical errors are superimposed to the best-fit MC (yellow histogram). Right: Background-model-dependent $2\nu\beta\beta$ fit. Double beta-like event rates in enriched-xenon (top) and depleted-xenon (bottom) superimposed to the best-fit MC, accounting for the radiogenic background contributions.}
	\label{fig:bg_fits}
\end{figure}

During the NEXT-White data taking, the background was measured using depleted xenon (<3$\%$ of $^{136}$Xe). In order to compute the half life of the two neutrino double beta decay, the NEXT technology offers the capability to perform a direct background subtraction by combining $^{136}$Xe-enriched and $^{136}$Xe-depleted data \cite{NEXT:2021dqj}. After this subtraction, an independent-background-model fit was performed achieving a $2\nu\beta\beta$ half life of $2.34 ^{+0.80}_{-0.46}(\textrm{stat}) ^{+0.30}_{-0.17}(\textrm{sys})\times 10^{21}$ years. The background-subtracted $2\nu\beta\beta$ event
energy spectrum is presented in Figure \ref{fig:bg_fits} left. To validate this result, a second fit of the event energy was implemented, background model dependent. There, the double beta candidates selected in the depleted and enriched xenon data taking, were jointly fitted to the radiogenic background model as shows Figure \ref{fig:bg_fits} right, providing a half life of $2.14 ^{+0.65}_{-0.38}(\textrm{stat}) ^{+0.46}_{-0.26}(\textrm{sys})\times 10^{21}$ years, demonstrating good agreement between the two fit strategies.

\section{NEXT-100}

The NEXT-100 detector started construction on 2021 in Spain and in USA and the assembly is planned for the first half of 2022 at the LSC. The second half of the year will be devoted to commissioning and calibration. The main goals comprise demonstrating the low background level predicted by the detailed Monte Carlo simulations performed, which consists of $5\times10^{-4}$ c/keV/kg/y, improve energy resolution to 0.5-0.7$\%$, background model assessment and prepare the tonne-scale \cite{NEXT:2015wlq}.


\smallskip

NEXT-100 will scale up NEXT-White by a factor between 2 and 3 in dimensions, will increase pressure from 10 bar to 15 bar and will incorporate a few improvements and changes concerning mainly the tracking plane. The SiPMs will be changed from the SensL series-C to the S13372-1350TE Hamamatsu with area 1.3 $\times$ 1.3 mm$^2$, collecting 60$\%$ more light. Additionally, the distance between SiPMs will be increased to 15 mm, in comparison with the 10 mm in the case of NEXT-White. The copper shield will be thicker and there won't be inner lead castle. As explained in \cite{NEXT:2015wlq}, NEXT-100 is expected to reach a sensitivity of about $6 \times 10^{25}$ yr after a run of 3 effective years.

\section{Tonne-scale R$\&$D}
The following step in the NEXT experiment is named NEXT-HD, the NEXT-tonne detector, whose main goal will be to increase the isotopic mass of Xe to explore the inverted hierarchy region of neutrino masses with a high-pressure xenon gas TPC. Backgrounds will be significantly reduced by replacing PMTs with SiPMs and operating at low temperatures. In the previous detectors, the electronics and the ceramic components of the PMTs are considered the main background sources. Additionally, the dark noise of the SiPMs decreases at low temperatures. The tracking resolution will increase by using a denser tracking plane with smaller pitch and reducing diffusion with gas additives such as helium. Furthermore, new readouts are under study: the possibility of using wavelength shifter optical fibers coupled to photosensors for t$_0$ and calorimetry or fast cameras for tracking \cite{NEXT:2020amj}.

\smallskip

A second phase of the NEXT-tonne will incorporate barium tagging to reach a half life sensitivity of 10$^{28}$ years.
When xenon decays neutrinoless double beta decay, it gives rise to the barium atom plus the two electrons. If the daughter atom in the decay is detected in coincidence with the event energy measurement, it would constitute a positive evidence of the $0\nu\beta\beta$ process \cite{McDonald:2017izm}. The barium ion can be tagged with the use of single molecule fluorescence imaging \cite{Rivilla:2020cvm}.

\section{Conclusions}

An overview of the NEXT experiment has been reported emphasizing the results of the NEXT-White detector, which comprise the excellent energy resolution below 1 $\%$ at 2.6 MeV, the efficient topology differentiation between signal and background events, the detailed background model and the two different fits for the $2\nu\beta\beta$ lifetime. NEXT-100 is currently under construction with the refinement focused in the tracking plane with respect to NEXT-White and afterwards the tonne-scale detector will explore the entire inverted-hierarchy region of neutrino masses.

\section*{Acknowledgments}

The speaker acknowledges support from the MCIN/AEI/10.13039/501100011033 of Spain and ERDF A way of making Europe under grant RTI2018-095979.

\bibliographystyle{JHEP}
\bibliography{mybibfile}

\providecommand{\href}[2]{#2}\begingroup\raggedright\begin{thebibliography}{1}

\bibitem{NEXT:2015wlq}
{\scshape NEXT} collaboration, \emph{{Sensitivity of NEXT-100 to Neutrinoless
  Double Beta Decay}},
  \href{https://doi.org/10.1007/JHEP05(2016)159}{\emph{JHEP} {\bfseries 05}
  (2016) 159} [\href{https://arxiv.org/abs/1511.09246}{{\ttfamily
  1511.09246}}].

\bibitem{Fukugita:1986hr}
M.~Fukugita and T.~Yanagida, \emph{{Baryogenesis Without Grand Unification}},
  \href{https://doi.org/10.1016/0370-2693(86)91126-3}{\emph{Phys. Lett. B}
  {\bfseries 174} (1986) 45}.

\bibitem{NEXT:2018rgj}
{\scshape NEXT} collaboration, \emph{{The Next White (NEW) Detector}},
  \href{https://doi.org/10.1088/1748-0221/13/12/P12010}{\emph{JINST} {\bfseries
  13} (2018) P12010} [\href{https://arxiv.org/abs/1804.02409}{{\ttfamily
  1804.02409}}].

\bibitem{NEXT:2019qbo}
{\scshape NEXT} collaboration, \emph{{Energy calibration of the NEXT-White
  detector with 1\% resolution near Q$_{\beta \beta}$ of$^{136}$Xe}},
  \href{https://doi.org/10.1007/JHEP10(2019)230}{\emph{JHEP} {\bfseries 10}
  (2019) 230} [\href{https://arxiv.org/abs/1905.13110}{{\ttfamily
  1905.13110}}].

\bibitem{NEXT:2020try}
{\scshape NEXT} collaboration, \emph{{Boosting background suppression in the
  NEXT experiment through Richardson-Lucy deconvolution}},
  \href{https://doi.org/10.1007/JHEP07(2021)146}{\emph{JHEP} {\bfseries 21}
  (2020) 146} [\href{https://arxiv.org/abs/2102.11931}{{\ttfamily
  2102.11931}}].

\bibitem{NEXT:2021dqj}
{\scshape NEXT} collaboration, \emph{{Measurement of the ${}^{136}$Xe
  two-neutrino double beta decay half-life via direct background subtraction in
  NEXT}},  \href{https://arxiv.org/abs/2111.11091}{{\ttfamily 2111.11091}}.

\bibitem{NEXT:2020amj}
{\scshape NEXT} collaboration, \emph{{Sensitivity of a tonne-scale NEXT
  detector for neutrinoless double beta decay searches}},
  \href{https://doi.org/10.1007/JHEP08(2021)164}{\emph{JHEP} {\bfseries 2021}
  (2021) 164} [\href{https://arxiv.org/abs/2005.06467}{{\ttfamily
  2005.06467}}].

\bibitem{McDonald:2017izm}
A.D.~McDonald et~al., \emph{{Demonstration of Single Barium Ion Sensitivity for
  Neutrinoless Double Beta Decay using Single Molecule Fluorescence Imaging}},
  \href{https://doi.org/10.1103/PhysRevLett.120.132504}{\emph{Phys. Rev. Lett.}
  {\bfseries 120} (2018) 132504}
  [\href{https://arxiv.org/abs/1711.04782}{{\ttfamily 1711.04782}}].

\bibitem{Rivilla:2020cvm}
I.~Rivilla et~al., \emph{{Fluorescent bicolour sensor for low-background
  neutrinoless double $\beta$ decay experiments}},
  \href{https://doi.org/10.1038/s41586-020-2431-5}{\emph{Nature} {\bfseries
  583} (2020) 48}.

\end{thebibliography}\endgroup

\end{document}